\documentstyle[12pt,twoside,psfig]{article}
\begin{document}

\begin{center}
{\Large\bf Curvature driven acceleration : a utopia or a reality ? 
}
\\[15mm]
Sudipta Das, \footnote{E-mail:dassudiptadas@rediffmail.com}~~
Narayan Banerjee \footnote{E-mail: narayan@juphys.ernet.in}

{\em Relativity and Cosmology Research Centre,\\Department of Physics, Jadavpur
University,\\ Calcutta - 700 032, ~India.}\\
and \\

Naresh Dadhich \footnote{E-mail: nkd@iucaa.ernet.in}

{\em Inter University Centre for Astronomy and Astrophysics,\\
Post Bag 4, Ganeshkhind, Pune 411007, ~India.}\\[15mm]
\end{center}

\vspace{0.5cm}
\vspace{0.5cm}

\pagestyle{myheadings}
\newcommand{\be}{\begin{equation}}
\newcommand{\ee}{\end{equation}}
\newcommand{\bea}{\begin{eqnarray}}
\newcommand{\eea}{\end{eqnarray}}

\begin{abstract}
The present work shows that a combination of nonlinear contributions 
from the Ricci curvature in Einstein field equations can drive a late 
time acceleration of expansion of the universe. The transit from the 
decelerated to the accelerated phase of expansion takes place smoothly 
without having to resort to a study of asymptotic behaviour. This 
result emphasizes the need for thorough and critical examination 
of models with nonlinear contribution from the curvature.
\end{abstract}

\section{INTRODUCTION}
The search for a dark energy component, the driver of the present 
accelerated expansion of the universe, has gathered a huge momentum 
because the alleged acceleration is now believed to be a certainty, 
courtesy the WMAP data \cite{spergel}. As no single candidate enjoys 
a pronounced supremacy over the others as the dark energy component 
in terms of its being able to explain all the observational details 
as well as having a sound field theoretic support, any likely 
candidate deserves a careful scrutiny until a final unambiguous 
solution for the problem emerges. The cosmological constant $\Lambda$, 
a minimally coupled scalar field with a potential, Chaplygin gas or 
even a nonminimally coupled scalar field are amongst the most popular 
candidates ( see ref. \cite{sahni} for a comprehensive review). Recently an 
attempt along a slightly different direction is gaining more and more 
importance. This effort explores the possibility whether geometry in 
its own right could serve the purpose of explaining the present 
accelerated expansion. The idea actually stems from the fact that 
higher order modifications of the Ricci curvature $R$, in the form of 
$R^2$ or $R_{\mu\nu}R^{\mu\nu}$ etc. in the Einstein - Hilbert action
could generate an accelerated expansion in the very early 
universe \cite{kerner}.  As the curvature $R$ is expected to fall off 
with the evolution, it is an obvious question if inverse powers of 
$R$ in the action, which should become dominant during  the later stages, 
could drive a late time acceleration.
\par A substantial amount of work in this direction is already there 
in the literature. Capozziello et al. \cite{capoz} introduced an action 
where $R$ is replaced by $R^n$ and showed that it leads to an 
accelerated expansion, i.e, a negative value for the deceleration 
parameter $q$ for $n=-1$ and $n=\frac{3}{2}$. Carroll et al. 
\cite{carroll} used a combination of $R$ and $\frac{1}{R}$, and a 
conformally transformed version of theory, where the effect of the 
nonlinear contribution of the curvature is formally taken care of by 
a scalar field, could indeed generate a negative value for the 
deceleration parameter. Vollick also used this $1/R$ term in the 
action \cite{vollick} and the resulting field equations allowed an 
asymptotically exponential and hence accelerated expansion. The dynamical 
behaviour of $R^{n}$ gravity had been studied in detail by Carloni et.al 
\cite{carloni}. A remarkable result obtained by Nojiri and Odinstov 
\cite{nojiri} shows that it may indeed be possible to attain an inflation at 
an early stage and also a late surge of accelerated expansion from the same 
set of field equations if the modified Lagrangian has the form 
$\it{L} = R + R^m + R^{-n} $ where $m$ and $n$ are positive integers. 
However, the solutions obtained are piecewise, i.e, large and small values of 
the scalar curvature $R$, corresponding to early and late time behaviour of 
the model respectively, are treated separately. But this clearly hints 
towards a possibility that different modes of expansion at various stages of 
evolution could be accounted for by a curvature driven dynamics. Other 
interesting investigations such as that with an inverse $sinh(R)$ 
\cite{borow} or with $lnR$ terms \cite{odin} in the action are also 
there in the literature.
\par The question of stability \cite{dolgov} and other problems 
notwithstanding, these investigations surely open up an interesting 
possibility for the search of dark energy in the non-linear contributions 
of the scalar curvature in the field equations. However, in most of these 
investigations so far mentioned, the present acceleration comes either 
as an asymptotic solution of the field equations in the large cosmic 
time limit, or even as a permanent feature of the dynamics of the universe. 
But both the theoretical demand \cite{pt} as well as observations \cite{riess} 
( see also ref \cite{spergel} ) clearly indicate that the universe 
entered into its accelerated phase of expansion only very recently and 
had been decelerating for the major part of its evolution. So the deceleration parameter $q$ must have a signature flip from a positive to a negative value 
only in a recent past.
\par In the present work, we write down the field equations for a general 
Lagrangian $f(R)$ and investigate the behaviour of the model for two 
specific choices of $f(R)$, namely $f(R) = R - \frac{\mu^4}{R}$ and 
$f(R) = e^{-\frac{R}{6}}$. 
\par Although the field equations, a set of fourth order differential 
equations for the scale factor $a$, could not be completely solved 
analytically, the evolution of the `acceleration' of the universe could 
indeed be studied at one go, i.e, without having to resort to a 
piecewise solution. The results obtained are encouraging, both the 
examples show smooth transitions from the decelerated to the 
accelerated phase. In this work we virtually assume nothing regarding the 
relative strengths of different terms and let them compete in their own 
way, and still obtain the desired transition in the signature of the 
deceleration parameter $q$. This definitely provides a very strong support 
for the host of investigations on curvature driven acceleration, particularly 
those quoted in references [5, 6 and 7].
\par In the evolution equation, $q$ is expressed as a function of $H$, the 
Hubble parameter. This enables one to write an equation with only $q$ to 
solve for; as the only other variable remains is $H$ which becomes the 
argument. This method appears to be extremely useful, although it finds 
hardly any application in the literature. The only example noted by us 
is the one by Carroll et al. \cite{carr}, which, however, describes the 
nature only in an asymptotic limit.
\par In the next section the model with two examples are described and 
in the last section we include some discussion.
\noindent
\section{Curvature driven acceleration}
\par The relevant action is 
\be
\it A = \int [\frac{1}{16\pi G}f(R) + L_{m}]\sqrt{-g} d^4x ,
\ee
where the usual Einstein - Hilbert action is generalized by replacing $R$ 
with $f(R)$, which is an analytic function  of $R$, and $L_{m}$ is the Lagrangian for 
all the matter fields. A variation of this action with respect to the 
metric yields the field equations as 
\be
G_{\mu\nu} = R_{\mu\nu} - \frac{1}{2}Rg_{\mu\nu} = {T_{\mu\nu}}^{c}+ {T_{\mu\nu}}^{M} , 
\ee
where the choice of units $8\pi G = 1$ has been made. ${T_{\mu\nu}}^{M}$ 
represents the contribution from matter fields scaled by a factor of 
$\frac{1}{f'(R)}$ and ${T_{\mu\nu}}^{c}$ denotes that from the curvature 
to the effective stress energy tensor. ${T_{\mu\nu}}^{c}$ is actually given as
\be
{T_{\mu\nu}}^{c} = \frac{1}{f'(R)}[\frac{1}{2} g_{\mu\nu}{(f(R) - R f'(R))} + 
      {f'(R)}^{;\alpha\beta} (g_{\mu\alpha} g_{\nu\beta} - 
        g_{\mu\nu} g_{\alpha\beta})] .
\ee
A prime indicates differentiation with respect to Ricci scalar $R$.  
It deserves mention that we use a variation of (1) w.r.t. the metric tensor 
as in Einstein - Hilbert variational principle and not a Palatini variation 
where $A$ is varied w.r.t. both the metric and the affine connections. As 
the actual focus of the work is to scrutinize the role of geometry alone in 
driving an acceleration in the later stages, we shall work without any matter 
content, i.e, $L_{m} = 0$ leading to ${T_{\mu\nu}}^{M} = 0$. So for a 
spatially flat Robertson - Walker spacetime, where
\be
ds^2 = dt^2 - a^2(t) [ dr^2 + r^2 d\theta^2 + r^2 sin^2\theta d\phi^2] ,
\ee
the field equations (2) take the form ( see ref. \cite{capoz} ) 
\bea
3\frac{\dot{a}^2}{a^2} = \frac{1}{f'}[{ \frac{1}{2}( f - Rf') - 
                       3\frac{\dot{a}}{a}\dot{R} f''}] ,~~~~~~~~~~~~~~~~\\
2\frac{\ddot{a}}{a} + \frac{\dot{a}^2}{a^2} = -\frac{1}{f'}[{ 2\frac{\dot{a}}{a}
\dot{R} f'' + \ddot{R} f'' + \dot{R}^2 f'''  - \frac{1}{2}( f - Rf')}] .
\eea
Here $a$ is the scale factor and an overhead dot indicates differentiation 
w.r.t. the cosmic time $t$. If $f(R) = R$, the equation (2) and hence (5) and 
(6) take the usual form of vacuum Einstein field equations. It should be 
noted that the Ricci scalar $R$ is given by 
\be
R = - 6[ \frac{\ddot{a}}{a} + \frac{\dot{a}^2}{a^2}] ,
\ee
and already involves a second order time derivative of $a$. As equation (6) 
contains $\ddot{R}$, one actually has a system of fourth order differential 
equations. 
\par It deserves mention at this stage that if $R$ is a constant, then 
whatever form of $f(R)$ is chosen except $f(R) = R$, equations (5) and (6) 
represent a vacuum universe with a cosmological constant and hence yield 
a deSitter solution, i.e, an ever accelerating universe. Evidently we are not 
interseted in that, we are rather in search of a model which clearly shows 
a transition from a decelerated to an accelerated phase of expansion of the 
universe. As we are looking for a curvature driven acceleration at late time, 
and the curvature is expected to fall off with the evolution, we shall take a 
form of $f(R)$ which has a sector growing with the fall of $R$. We work out 
two examples where indeed the primary purpose is served. 
\\
\\
(i) $f(R) = R - \frac{\mu^4}{R}.$
\\
\\
\par In the first example, we take 
\be
f(R) = R - \frac{\mu^4}{R} ,
\ee
where $\mu$ is a constant. Indeed $\mu$ has a dimension, that of 
$R^\frac{1}{2}$, i.e, that of $(time)^{-1}$.  This is exactly the form 
used by Carroll et al. \cite{carroll} and Vollick \cite{vollick}. 
Using the expression (8) in a 
combination of the field equations (5) and (6), one can easily arrive at the 
equation 
\be
2\dot{H} = \frac{1}{(R^2 + \mu^4)}[{ 2\mu^4 \frac{\ddot{R}}{R} - 
6\mu^4\frac{\dot{R}^2}{R^2} - 2\mu^4 H \frac{\dot{R}}{R}}] , 
\ee
where $H = \frac{\dot{a}}{a}$, is the Hubble parameter. As both $R$ and $H$ 
are functions of $a$ and its derivatives, equation (9) looks set for 
yielding the solution for the scale factor. But it involves fourth order 
derivatives of $a$ ( $R$ already contains $\ddot{a}$ ) and is highly 
nonlinear. This makes it difficult to obtain a completely analytic solution 
for $a$. As opposed to the earlier investigations where either a piecewise 
or an asymptotic solution was studied, we adopt the following strategy. 
The point of interest is the evolution of the deceleration parameter 
\be
q = -\frac{a \ddot{a}}{\dot{a}^2} = -\frac{\dot{H}}{H^2} - 1 . 
\ee
So we translate equation (9) into the evolution equation for $q$ using 
equation (10) and obtain \\

$~~~~~~~~~~~~~~\mu^4\frac{\ddot{q}}{(q - 1)} - 3\mu^4 \frac{\dot{q}^2}{(q - 1)^2} 
+ \mu^4 H^2 (q + 1)(4q + 7) - 3\mu^4 q (q + 1) H^2 \frac{(2q - 3)}{(q - 1)}$ 
\be
- 10 \mu^4 H^2 (q + 1)^2 + 36 H^6 (q - 1)^2 (q + 1) = 0 .
\ee
This equation, although still highly nonlinear, is a second order equation 
in $q$. But the problem is that both $q$ and $H$ are functions of time and 
cannot be solved for with the help of a single equation. However, they are not 
independent and are connected by equation (10). So we replace time derivatives 
by derivatives w.r.t. $H$ using equation (10) and write (11) as 
\be 
\frac{1}{3}(q^2 - 1)H^2 q^{\dagger\dagger} + {[\frac{1}{3} (q - 1) + (q + 1)]} 
H^2{ q^{\dagger}}^2 + \frac{2}{3}H ( q^2 - 1) q^{\dagger} + H^4 ( q - 1)^4 
- ( q - 1)( 4q^2 - 4q -1) = 0 . 
\ee
Here for the sake of simplicity $\mu^4$ is chosen to be 12 ( in proper units ), and a dagger represents a differentiation 
w.r.t. the Hubble parameter $H$. As $\frac{1}{H}$ 
is a measure of the age of the universe and $H$ is a monotonically decreasing 
function of the cosmic time, equation (12) can now be used as the evolution equation for $q$. The equation appears to be hopelessly nonlinear to give an 
analytic solution but if one provides two initial conditions, for $q$ and 
$q^{\dagger}$, for some value of $H$, a numerical solution is definitely 
on cards. We choose units so that $H_{0}$, the present value of $H$, is 
unity and pick up sets of values for $q$ and $q^{\dagger}$ for $H = 1$ 
( i.e, the present values ) from observationally consistent region \cite{alam} 
and plot $q$ versus $H$ numerically. As the inverse of $H$ is the estimate 
for the cosmic age, `future' is given by $H < 1$ and past by $H > 1$. 
The plots speak for themselves. One has the desired feature of a 
negative $q$ at $H = 1$ and it comes to this negative phase only in the 
recent past. Furthermore, in near future, $q$ has another sign flip in 
the opposite direction, forcing an exhibition of a decelerated expansion 
in the future again. An important point to note here is that 
neither the nature 
of the plots, nor the values of $H$ at which the transitions take place, 
crucially depend on the choice of initial conditions, so the model is 
reasonably stable. It is interesting to note the asymptotic values of 
$q$. For very high values of the cosmic time, i.e, for $H \rightarrow 0$ ,
$q \rightarrow 1$ or $q \rightarrow (\frac{1 \pm \sqrt{2}}{2})$. So it admits 
at least one negative value, consistent with the results obtained in ref. [5]. 
As $f(R)$ contains $R^n$ with $n \leq 1$, one does not expect this to give 
rise to an early inflation. For $H \rightarrow \infty$, i.e, for 
$t \rightarrow 0$, $q \rightarrow +1$, which exhibits indeed a non-inflationary 
decelerating model.
\\
\begin{figure}[!h]
\mbox{\psfig{figure=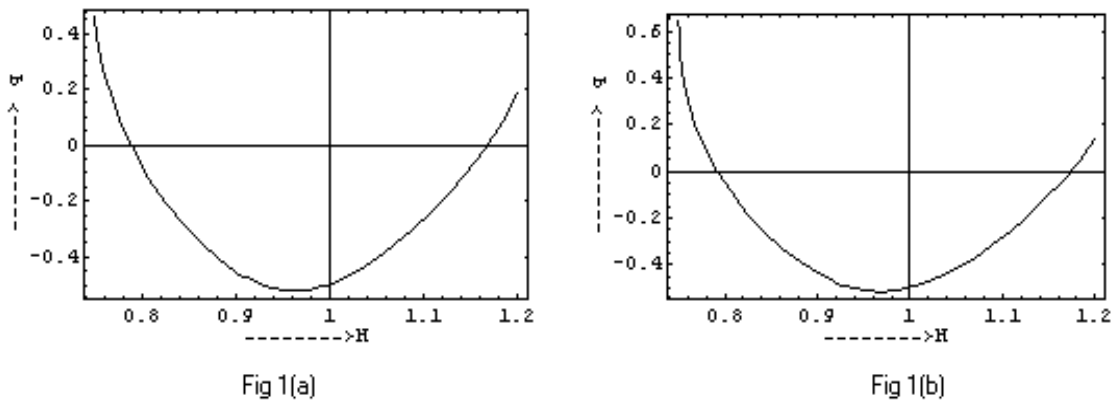,height=2.3in,width=6.0in}}
\caption{Figure 1(a) and 1(b) shows the plot of $q$ vs. $H$ for
$f(R) = R - \frac{\mu^4}{R}$ for different initial conditions. For figure 1(a) 
we choose the initial conditions as $q[1] = -0.5$, $q'[1] = 1.2$ whereas for 
figure 1(b) we set the initial conditions as $q[1] = -0.5$, $q'[1] = 1.0$.}
\label{first_fig}
\end{figure}
\\
\\
(ii)  $f(R) = e^{-\frac{R}{6}}$ .
\\
\par In this choice, the function $f$ is monotonically increasing with $t$ 
as $R$ is decreasing with $t$. 
The field equations (5) and (6) have the form 
\bea
3\frac{\dot{a}^2}{a^2} = -6[{ \frac{1}{2}(1 + \frac{R}{6}) - 
              \frac{1}{12}\frac{\dot{a}}{a}\dot{R} }] ,~~~~~~~~~~~~~~~~\\
2\frac{\ddot{a}}{a} + \frac{\dot{a}^2}{a^2} = -6[{\frac{1}{2}(1 + \frac{R}{6}) 
                        - \frac{1}{18}\frac{\dot{a}}{a}\dot{R} - 
                  \frac{1}{36}\ddot{R} + \frac{1}{216}\dot{R}^2 }] .
\eea
From these two equations it is easy to write 
\be
2\dot{H} = \frac{1}{6}\ddot{R} - \frac{1}{36}\dot{R}^2 - \frac{1}{6}H \dot{R} .
\ee
 Following the same method as before, the evolution of $q$ as a function of 
$H$ can be written as
\\

$H^4 (q + 1)q^{\dagger\dagger} + { [H^4 - (q + 1)H^6}] {q^{\dagger}}^2 
+ [ 2(q + 1)H^3 + 6qH^3 + 3H^3 - 4H^5( q^2 - 1)]q^{\dagger}$
\be 
+ 6 H^2 q^2 + 2 H^2 q - 4H^4 (q^2 - 1)(q - 1) - 8H^2 + 2 = 0 .
\ee

With similar initial conditions for $q$ and $q^{\dagger}$ at $H = 1$, 
the plot of $q$ versus $H$ ( figure 2 ) shows features very similar to 
the previous example, the deceleration parameter $q$ has two signature 
changes, from a positive to a negative phase in the recent past ( $H > 1$ ) 
and in the reverse direction in the near future ( $H < 1$ ). In this case also,  a small change in initial conditions hardly has any perceptible change 
in the graphs. 
\par As for the asymptotic behaviour, one expects that as exponential 
functions have higher powers, the model could lead to an early inflation. 
Actually it does; for $H \rightarrow \infty$, i.e, for small $t$, 
$q \rightarrow -1$. The equation (16) does not straightaway lead to the 
other extreme end, $H \rightarrow 0$, i.e, $t \rightarrow$ large. Actually 
a common factor of $(q + 1)$ had been cancelled while arriving at equation 
(16). So $q = -1$ is the other solution, which holds for $H \rightarrow 0$ 
limit. So here also, the model has a steady acceleration at the far 
future end. 
\begin{figure}[!h]
\mbox{\psfig{figure=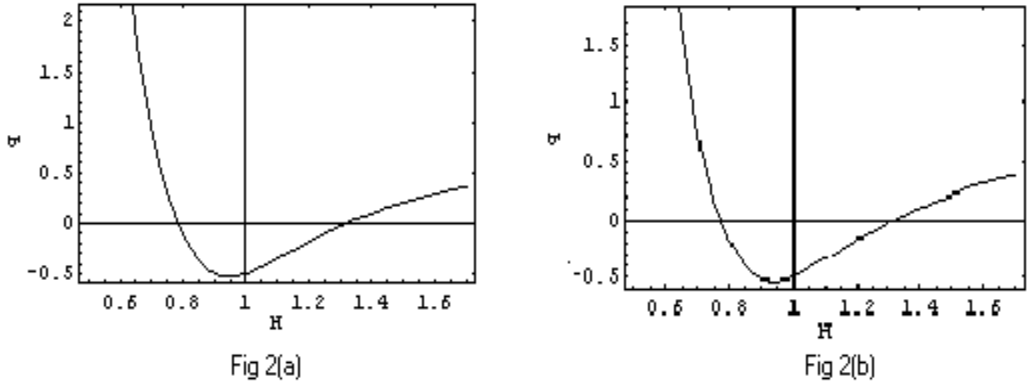,height=2.3in,width=5.7in}}
\caption{Figure 2(a) and 2(b) shows the plot of $q$ vs. $H$ for
$f(R) = e^{-\frac{R}{6}}$ for different initial conditions. Here also for 
figure 2(a) and 2(b) we set the initial conditions as $q[1] = -0.5$, 
$q'[1] = 1.0$ and  $q[1] = -0.5$, $q'[1] = 1.2$ respectively.}
\label{second_fig}
\end{figure}

\par As the plots provide a sufficient data set, attempts could be made 
to find the closest analytical expression for $ q = q(H)$. These 
expressions are found to be polynomials. For example, a very close 
analytical expression for figure 2(b), within the accuracy of plots, is 
given as 
\be
  q = 47.95 H^6 - 335.73 H^5 + 991 H^4 - 1586.90 H^3 + 1459.20 H^2 
 - 729.73 H + 153.74  . 
\ee
This expression holds only when $H$ is reasonably close to one, and has 
nothing to do with other ranges of values of $H$.
\section{Discussion}
\par The present work indicates that by asking the question whether geometry 
in its own right can lead to the late surge of accelerated expansion, 
some feats can surely be achieved. Both the examples considered here 
indicate that one can build up models which starts accelerating at the 
later stage of evolution and thus allow all the past glories of the 
decelerated model like nucleosynthesis or structure formation to remain 
intact. An added bonus of the examples is that in both the cases the 
universe re-enters a decelerated phase in near future and the `phantom 
menace' is avoided - the universe does not have to have a singularity 
of infinite volume and infinite rate of expansion in a `finite' future.
\par It is of course true that a lot of other criteria has to be satisfied 
before one makes a final choice, and we are nowhere near that. Already 
there is a criticism of $\frac{1}{R}$ gravity that it is unsuitable 
for local astrophysics because of problems regarding stability \cite{dolgov}.
However, it was pointed out by Nojiri and Odinstov \cite{nojiri} that a 
polynomial may save the situation ( see also reference \cite{barrow} ). 
Our second example is exponential in $R$, 
i.e, a series of positive powers in $R$, and hence could well satisfy the 
criterion of stability. As already pointed out, although the choice of 
$f(R) = R - \frac{\mu^4}{R}$ is already there in the literature and 
served the purpose in a restricted sense than it does in the present work, 
the choice of $f(R) = e^{-\frac{R}{6}}$ has hardly any mention in the 
literature.
\par It should also be noted that the present toy model deals with a
 vacuum universe and one has to either put in matter, or derive the relevant 
matter at the right epoch from the curvature itself. Some efforts 
towards this have already begun \cite{odinstov}.  
On the whole, there are reasons to be optimistic about a curvature driven 
acceleration which might become more and more important in view 
of the fact that WMAP data could indicate a very strong constraint on the 
variation of the equation of state parameter $w$ \cite{paddy}.
\vskip .2in
\section{Acknowledgement}
Authors are thankful to Mriganka Chakraborty for useful discussion.


\vskip .2in

\begin{thebibliography}{25}

\bibitem{spergel} D. N. Spergel et al., Astrophys. J. Suppl., {\bf 148}, 175 (2003);\\
                  L. Page, astro-ph/0302220;\\
                  L. Verde et al., Astrophys. J. Suppl., {\bf 148}, 195 (2003);\\
                  S. Bridle, O. Lahav, J. P. Ostriker and P. J. Steinhardt,
                  Science, {\bf 299}, 1532 (2003);\\
                  C. Bennet et al., astro-ph/0302207;\\
                  G. Hinshaw et al., astro-ph/0302217;\\
                  A. Kognt et al., astro-ph/0302213.
\bibitem{sahni}  V. Sahni, astro-ph/0403324;\\
                 T. Padmanabhan, Phys. Rep., {\bf 380}, 235 (2003).   
\bibitem{kerner} A. A. Starobinsky, Phys. Lett. B, {\bf 91}, 99 (1980);\\
                 R. Kerner, Gen. Rel. Gravit., {\bf 14}, 453 (1982);\\
                 J. P. Duruisseau, R. Kerner, Class. Quant. Grav., 
                     {\bf 3}, 817 (1986).    
\bibitem{capoz}  S. Capozziello, S. Carloni, A. Troisi, astro-ph/0303041;\\
                 S. Capozziello, V. F. Cardone, S. Carloni, A. Troisi, 
                     astro-ph/0307018.
\bibitem{carroll} S. M. Carroll, V. Duvvuri, M. Trodden, M. S. Turner, 
                      astro-ph/0306438.
\bibitem{vollick} D. N. Vollick, Phys. Rev. D, {\bf 68}, 063510 (2003).
\bibitem{carloni} S. Carloni, P.K.S. Dunsby, S. Capozziello and A. Troisi,
                     gr-qc/0410046.
\bibitem{nojiri} S. Nojiri and S. D. Odinstov, Phys. Rev. D, 
                      {\bf 68}, 123512 (2003).
\bibitem{borow} A. Borowiec and M. Francaviglia, hep-th/0403264.
\bibitem{odin} S. Nojiri and S. D. Odinstov, hep-th/0308176.
\bibitem{dolgov} A. D. Dolgov and M. Kawasaki, astro-ph/0307285.
\bibitem{pt} T. Padmanabhan and T. Roychoudhury, astro-ph/0212573;\\
             T. Roychoudhury and T. Padmanabhan, Astron. Astrophys, 
              {\bf 429}, 807 (2005).
\bibitem{riess} A. G. Riess, astro-ph/0104455.
\bibitem{carr} S. M. Carroll et al., astro-ph/0410031.
\bibitem{alam} U. Alam, V. Sahni, T. D. Saini and A. A. Starobinsky,\\ 
               ~~~~~~~~~~~~~~~~~~~~~~~ Mon. Not. Roy. Ast. Soc., {\bf 344}, 
                                       1057 (2003); 
                   [ astro-ph/0303009];\\
               U. Alam and V. Sahni, astro-ph/0209443.
\bibitem{barrow} J. D. Barrow and A. C. Ottewill, J. Phys. A, {\bf 16}, 
                       2756 (1983);\\
                 S. Nojiri and S. D. Odinstov, Mod. Phys. Lett. A, {\bf 19}, 
                  627 (2004).   
\bibitem{odinstov} M. Abdalla, S. Nojiri and S. D. Odinstov, Class. Quant. 
                    Grav., {\bf 22}, L35 (2005);\\
                   G. Allemandi, A. Borowiec, M. Francaviglia and 
                    S. D. Odinstov, gr-qc/0504057.
\bibitem{paddy} H. K. Jassal, J. S. Bagla and T. Padmanabhan, 
                 Mon. Not. R. Astron. Soc., {\bf 356}, L11 (2005).     

 
\end{thebibliography}
\end{document}